\documentclass[%
reprint,
superscriptaddress,
amsmath,amssymb,
aps,
prl,
]{revtex4-2}

\usepackage{graphicx}
\usepackage{dcolumn}
\usepackage{bm}
\usepackage{gensymb}
\usepackage{subfloat}
\usepackage{xcolor}
\usepackage{comment}
\usepackage{epstopdf}
\usepackage{multirow}
\usepackage[normalem]{ulem}
\usepackage{subfigure}
\usepackage{hyperref}

\usepackage{xcolor}
\definecolor{Blue}{rgb}{0,0,1}
\definecolor{Red}{rgb}{1,0,0}
\definecolor{Green}{rgb}{0,0.52,0.0}
\definecolor{orange}{rgb}{1,0.5,0}
\definecolor{orange2}{rgb}{1,0.5,0.5}

\begin{document}
	
	\preprint{APS/123-QED}
	
	\title{Theoretical issues in the accurate computation of the 
	electron-phonon interaction contribution to the total energy}%
	
	\author{Shilpa Paul}%
	\affiliation{%
		Department of Metallurgical Engineering and Materials Science, Indian Institute of Technology Bombay, Mumbai 400076, India
	}%
	
	\author{M. P. Gururajan}%
	\affiliation{%
		Department of Metallurgical Engineering and Materials Science, Indian Institute of Technology Bombay, Mumbai 400076, India
	}%

	\author{Amrita Bhattacharya}%
	\affiliation{%
		Department of Metallurgical Engineering and Materials Science, Indian Institute of Technology Bombay, Mumbai 400076, India
	}%
	\author{T. R. S. Prasanna}%
	\email{prasanna@iitb.ac.in}
	\affiliation{%
		Department of Metallurgical Engineering and Materials Science, Indian Institute of Technology Bombay, Mumbai 400076, India
	}%
	
	\date{\today}
	
\begin{abstract}
We report the computation of the standard Hamiltonian of a coupled 
electron-phonon system by accurately computing the electron-phonon interaction 
(EPI) contribution to the total energy. This gives the most accurate \textit{ab 
initio} total energy till date. However, our results show that the per-atom EPI 
energy is unit-cell-size dependent due to the partial-Fan-Migdal term that 
arises from the anti-symmetric nature of the crystal wavefunction. Due to this, 
only energy differences between polytypes, in supercells with identical number 
of atoms, are meaningful, rather than per-atom total energy. This violates our 
understanding of Quantum Mechanics applied to periodic solids and raises 
serious theoretical questions. In his original (1951) paper, Fan suggested, 
without specifying any reason, that second-order perturbation theory applied to 
the whole crystal is of questionable validity. Our results support Fan’s 
suggestion for the reason that the partial-FM term makes the per-atom total 
energy unit-cell-size dependent. This leads to a new fundamental problem in 
condensed matter physics, viz. second-order perturbation theory is invalid for 
whole crystals, an entire class. It is essential to resolve this problem, 
especially because it causes the standard Hamiltonian, the starting point of 
EPI studies and the most accurate  \textit{ab initio} total energy, to be of 
questionable validity.
\end{abstract}
	
	\maketitle
	
	
	The total energy is of fundamental importance in \textit{ab 
	initio} studies and is the basis of the density functional theory (DFT) 
	\cite{Kohn1964, Kohn1965}. For crystalline solids, because of the 
	connection between Quantum Mechanics and lattice periodicity through the 
	periodic potential, U(\textbf{r}),  \cite{Ashcroft1976solid, 
	Martin2020electronic} the accepted norm is to determine the per-atom total 
	energy and other properties only for the primitive unit cell (PC). These 
	quantities are identical for the PC and supercells (SC) and are independent 
	of unit cell size. Consequently, in 
	\textit{ab initio} studies, the total energy is first minimized, using 
	variational methods, with respect to structural parameters for the PC. 
	Subsequently, the band-structure, electronic, optical properties etc. are 
	computed for the relaxed PC.

	The \textit{ab initio} total energy is the primary quantity that is 
	computed to determine the stable structure and the relative energy 
	differences of the metastable structures. Further, there is currently great 
	interest in crystal structure prediction, especially, for new low-lying 
	metastable structures of various materials. These studies mainly involve 
	search for new metastable structures using a variety of techniques and 
	their energy differences with respect to the stable structure 
	\cite{Oganov2019structure, Catlow2020structure, Cheng2022crystal, 
	Hong2020training, Yamashita2018crystal}. In such studies, the total 
	energy differences at 0 K are computed using DFT. 

	DFT studies \cite{Kohn1964, Kohn1965} have been successful in determining 
	the stable structure and energy differences between polymorphs for many 
	materials. However, for some materials, the DFT and experimental results 
	vary. Hence, additional contributions from van der Waals (vdW) dispersion 
	(computed as part of the DFT computation \cite{Kawanishi2016, Scalise2019, 
	Ramakers, Cazorla2019})	and zero-point vibrational energy (ZPVE) 
	(computed separately from DFPT or FDM methods) are also included to obtain 
	(more) accurate total energy \cite{Kawanishi2016, Scalise2019, Ramakers, 
	Cazorla2019, Van2007, Bhat2018}. 

	Clearly, it is necessary to compute the \textit{ab initio} total energy as 
	accurately as possible. In this regard, we have recently proposed the 
	inclusion of electron-phonon interaction (EPI) contribution to the total 
	energy \cite{Varma2022}. Referring to our work, Allen has formally derived 
	the expression for including the same from second-order perturbation theory 
	\cite{Allen2020, Allen2022erratum}. This equation, as such, cannot be 
	computed using available software. Hence, in earlier studies, we have 
	approximated Allen's equation \cite{Allen2022erratum} and computed the EPI 
	contribution to the total energy for C, Si, SiC and BN polymorphs 
	\cite{Varma2022, Paul2022}.
	

	However, for accurate total energy studies, it is necessary to compute the 
	complete Allen's equation (Eq. 4 of Ref. \cite{Allen2022erratum}) for the 
	EPI contribution. In this paper, we recast Allen's equation for 
	the case of semiconductors and insulators. The recast (equivalent) 
	expression can be computed using existing software, leading to more 
	accurate total energy for all semiconductors and insulators. 

	It is important to place the recent EPI studies \cite{Varma2022, 
	Allen2022erratum} in perspective. The standard Hamiltonian 
	\cite{Allen-chakraborty1978theory, Giustino2017, Allen2022erratum} for a 
	coupled electron-phonon system is given by \cite{Allen2022erratum} 
\begin{equation}
H_{\mathrm{eff}}= \sum_k \epsilon_k(V) c_k^{\dagger} c_k+\sum_Q \hbar 
\omega_Q\left(a_Q^{\dagger} a_Q+1 / 2\right)+V_{\mathrm{ep}} 
\label{Std-Hamiltonian-EPI}
\end{equation}
\begin{equation}
\begin{aligned}
V_{\mathrm{ep}}= &  \sum_{k Q} V^{(1)}(k Q) c_{k+Q}^{\dagger} 
c_k\left(a_Q+a_{-Q}^{\dagger}\right) \\
& +\sum_{k Q Q^{\prime}} V^{(2)}\left(k Q Q^{\prime}\right) 
c_{k+Q+Q^{\prime}}^{\dagger} c_k
\\ &   \mathrm{x} 
\left(a_Q+a_{-Q}^{\dagger}\right)\left(a_{Q^{\prime}}+a_{-Q^{\prime}}^{\dagger}\right)
+ \ldots
\label{Vep}
\end{aligned}
\end{equation}
	Eq. \ref{Std-Hamiltonian-EPI} gives the \textit{total energy} of a coupled 
	electron-phonon system. The first two terms give the DFT total energy and 
	the ZPVE contribution respectively \cite{Allen2022erratum,Giustino2022ab}. 
	Crucially, the $V_{ep}$ term has never been computed till date 
	because the emphasis has been to compute the EPI contributions to 
	individual eigenstates and not to the whole crystal. Even the expression 
	for the $V_{ep}$ term has only recently been derived by Allen 
	\cite{Allen2022erratum} referring to our earlier work \cite{Varma2022}. In 
	this paper we compute the $V_{ep}$ term for the first time. \textit{Thus, 
	for the first time, we report the computation of the standard Hamiltonian 
	for a coupled electron-phonon system.} It is evident that the standard 
	Hamiltonian gives the most accurate \textit{ab initio} total energy. Thus, 
	the present work will be of great importance in accurately determining the 
	energy differences between polymorphs, for the applications discussed above 
	\cite{Oganov2019structure, Catlow2020structure, Cheng2022crystal, 
	Hong2020training, Yamashita2018crystal, Kawanishi2016, Scalise2019, 
	Ramakers, Cazorla2019, Van2007, Bhat2018}.

	However, our results raise serious theoretical issues. Our results show that 
	the partial-Fan-Migdal term in Allen's equation makes the per-atom EPI 
	(and total) energy unit-cell size dependent. This term arises due to the 
	anti-symmetric nature of the crystal wavefunction or the Pauli principle. 
	This is likely the first instance where the energy per-atom depends on unit 
	cell size, even when the lattice remains periodic. This violates our 
	understanding of Quantum Mechanics applied to periodic solids  
	\cite{Ashcroft1976solid, Martin2020electronic}. 

	Importantly, in the discussion of the partial-FM term in his original 
	(1951) paper, Fan \cite{Fan1951} suggested, without specifying any reason, 
	that second-order perturbation theory applied to the whole crystal is of 
	questionable validity. Our results 	support Fan's suggestion for the reason 
	that the partial-FM term causes the per-atom energy to be 	
	unit-cell-size dependent. In continuation of his concerns, Fan 
	\cite{Fan1951} proposed an approximation, viz. the EPI(FM) 
	contribution to the band-structure energy is the EPI contribution to the 
	total energy. However, Allen \cite{Allen2022EPIBS} has shown that this 
	approximation (made independently  in our earlier work \cite{Varma2022}) is 
	theoretically invalid. Currently, neither the exact nor the approximate 
	expressions for the EPI contribution to the total energy are theoretically 
	satisfactory. 

	Thus, for the first time in a century of Quantum Mechanics, 
	it appears that second-order perturbation theory is invalid for an entire 
	class of systems, whole crystals. \textit{This can be considered to be a 
	new fundamental problem in condensed matter physics.} It is essential to 
	resolve this problem, especially because it causes the standard 
	Hamiltonian, Eq. \ref{Std-Hamiltonian-EPI}, the starting point 
	of EPI studies \cite{Giustino2017} and also the most accurate \textit{ab 
	initio} total energy of materials, to be on questionable theoretical 
	foundations.

	In the Allen-Heine theory, EPI contribution to the eigenenergy 
	arises from the Debye-Waller (DW) and Fan-Migdal (FM) terms and, at 0 K, is 
	given by  
	\cite{Allen1976, Allen1980}
	\begin{equation}
	\Delta \epsilon_{n\textbf{k}}(0) = 
	\Delta\epsilon_{n\textbf{k}}^{\textrm{DW}}(0) + 
	\Delta\epsilon_{n\textbf{k}}^{\textrm{FM}}(0)
	\label{EPI-Energy-eigenstate} 
	\end{equation}

	The FM self-energy term includes contributions from both conduction and 
	valence bands \cite{Allen1976, Allen1980}. It can also be written as a sum 
	of their separate contributions as \cite{Nery2018} 
\begin{equation}
	\Delta\epsilon_{n\textbf{k}}(0) = 
		\Delta\epsilon_{n\textbf{k}}^{\textrm{FM},\textrm{occ}}(0)
		+\Delta\epsilon_{n\textbf{k}}^{\textrm{FM},\textrm{unocc}}(0) 
	\label{FM-occ-unocc} 
	\end{equation}

These partial-FM self-energy terms are given by \cite{Nery2018}
\begin{equation}
	\Delta \epsilon_{n\textbf{k}}^{\textrm{FM,unocc}}(0) = \frac{1}{N_\textbf{q}} 
	\sum^{\textrm{BZ}}_{\textbf qj} \sum^{\textrm{unocc}}_{n'} 
	\frac{{|\langle\textbf{k}n|H_j^{(1)}|\textbf{k}+\textbf{q}n'\rangle|}^2}{\omega
	 - \epsilon_{\textbf{k}+\textbf{q}n'}-\omega_{\textbf qj}+i\eta'}
	\label{FM-unocc}
\end{equation}

and 
\begin{equation}
	\Delta \epsilon_{n\textbf{k}}^{\textrm{FM,occ}}(0) = \frac{1}{N_\textbf{q}} 
	\sum^{\textrm{BZ}}_{\textbf qj} \sum^{\textrm{occ}}_{n'} 
	\frac{{|\langle\textbf{k}n|H_j^{(1)}|\textbf{k}+\textbf{q}n'\rangle|}^2}{\omega
	 - \epsilon_{\textbf{k}+\textbf{q}n'}+\omega_{\textbf qj}+i\eta'}
	\label{FM-occ}
\end{equation}

The EPI alters all eigenenergies and hence, the total energy is also 
altered \cite{Varma2022}. Allen has derived the expression for the EPI 
contribution to total energy, $\Delta E_{\textrm{EP}} (V,0)$, from  
second-order perturbation theory, as \cite{Allen2022erratum, Allen2023}
\begin{multline}
	\Delta E_{\textrm{EP}} (V,0) = \\
\sum_k \left[\langle k 
|V^{(2)}|k\rangle + \sum_Q \frac{|\langle k 
|V^{(1)}|k+Q\rangle|^2}{\epsilon_k - 
\epsilon_{k+Q}-\hbar\omega_{-Q}}(1-f_{k+Q}) \right]\;f_k
	\label{EPI-zero-K-Allen}
\end{multline}
	
	The structure of Eq. \ref{EPI-zero-K-Allen} is similar to the sum over 
	occupied states of the EPI contribution to individual eigenenergy, $\Delta 
	\epsilon_{n\textbf{k}}(0)$ \cite{Allen1980}, and differs in the 
	presence of the ($1-f_{k+Q}$) factor in the FM (second) term. 
	\cite{Allen2022erratum}. (The first term is the DW term.) (An explicit 
	version for the FM term in Eq. \ref{EPI-zero-K-Allen} is 
	given in the Supp. Info. I). This FM term is identical to that in the 
	earliest EPI studies of Fr{\"o}hlich (Eq.2.11) \cite{Frohlich1950theory} 
	and Fan (Eq.4) \cite{Fan1951}. The presence of the ($1-f_{k+Q}$) factor is 
	to ensure that the $k+Q$ is an empty state to satisfy the Pauli principle 
	\cite{Frohlich1950theory, Fan1951}. 

	The existing computational packages cannot compute  Eq. 
	\ref{EPI-zero-K-Allen} due to the presence of the ($1-f_{k+Q}$) factor.
	
	We describe below a method to accurately compute  Eq. 
	\ref{EPI-zero-K-Allen} based on the EPI implementation (TDepES) in the 
	ABINIT software package \cite{Gonze2009, Gonze2016, Abinit}. This module 
	has been widely used to obtain temperature dependent (due to EPI) 
	band-structures and band-gaps \cite{Ponce2015, Ponce2014b, Ponce2014a, 
	Antonius2015, Friedrich2015, Tutchton2018, Querales2019, Miglio2020}. 
 
	For semiconductors and insulators, the Fermi-Dirac distribution function, 
	$f_{k+Q}$ and $f_k$, equals 1 for the occupied (valence) states and 0 
	for the unoccupied (conduction) states \cite{Nery2018}. Thus, only 
	unoccupied bands contribute to the second term in Eq. 
	\ref{EPI-zero-K-Allen}  (Supp Info I, \cite{Fan1951}). Equivalently, it 
	becomes $\Delta\epsilon_{n\textbf{k}}^{\textrm{FM}}(0) - 
	\Delta\epsilon_{n\textbf{k}}^{\textrm{FM,occ}}(0)$, by opening 
	the ($1-f_{k+Q}$) factor. Thus, Eq. \ref{EPI-zero-K-Allen} can be rewritten 
	as
\begin{equation}
\begin{split}
	\Delta E_{\textrm{EP}} (V,0) &= 
\sum^{\textrm{occ}}_{n,\textbf{k}} \left[ 
\Delta\epsilon_{n\textbf{k}}^{\textrm{DW}}(0) + 	
\Delta\epsilon_{n\textbf{k}}^{\textrm{FM,unocc}}(0)\right] \\ &=
\sum^{\textrm{occ}}_{n,\textbf{k}} \left[ \Delta \epsilon_{n\textbf{k}}(0) - 	
\Delta\epsilon_{n\textbf{k}}^{\textrm{FM,occ}}(0)\right] \\
&= \Delta E^{\textrm{ep}}_{\textrm{bs}}(V,0) - \Delta 
E_{\textrm{tot}}^{\textrm{FM,occ}}(V,0)
	\label{EPI-zero-K-Allen-rewritten}
\end{split}
\end{equation}

	The equivalent expression, Eq. \ref{EPI-zero-K-Allen-rewritten}, shows that 
	the EPI contribution to the total energy comes from the EPI contribution to 
	the band-structure energy (sum over EPI contribution to the occupied 
	eigenstates), $\Delta E^{\textrm{bs}}_{\textrm{ep}}(V,0)$, 
	corrected by the total FM-occ contribution, $\Delta 
	E_{\textrm{tot}}^{\textrm{FM,occ}}$ (the partial-FM-occ term summed over 
	occupied states). The partial-FM-occ term is present due to the
	anti-symmetric nature of the crystal wavefunction and is absent for EPI 
	contribution to individual eigenstates 	\cite{Allen1980}. Its serious 
	implication will be seen below.

 	To compute Eq. \ref{EPI-zero-K-Allen-rewritten}, we first run 
 	the TDepES module \cite{Abinit} with the optimized number of bands, $M = 
 	N_{\textrm{tot}}$, that includes conduction and valence bands \cite{Ponce2015}. 
 	This will give the EPI correction to the eigenenergy, 
 	$\Delta\epsilon_{n\textbf{k}}(0)$, the first term in Eq. 
 	\ref{EPI-zero-K-Allen-rewritten}. We run the TDepES module a second time 
 	with $M = N_{\textrm{val}}$ to obtain the second term in Eq. 
 	\ref{EPI-zero-K-Allen-rewritten}, 	
 	$\Delta\epsilon_{n\textbf{k}}^{\textrm{FM,occ}}(0)$, the partial-FM-occ 
 	term.  Eq. \ref{EPI-zero-K-Allen-rewritten} can be rewritten as 
	a weighted sum over the k-points in the Irreducible Brillouin Zone (IBZ). 
	Thus, the EPI contribution to the total energy is computed as 
	\begin{multline}
	\Delta E_{\textrm{EP}} (V,0) = \\
2 \sum_{n,\textbf{k}}^{\textrm{occ,IBZ}} 
		w_\textbf{k}\Delta \epsilon_{n\textbf{k}} 
	(V,0) - 2 \sum_{n,\textbf{k}}^{\textrm{occ,IBZ}} 
	w_\textbf{k}\Delta\epsilon_{n\textbf{k}}^{\textrm{FM,occ}}	(V,0) 
	\label{EPI-accurate-Abinit} 
	\end{multline}

	\textit{Using Eq. \ref{EPI-accurate-Abinit} to include EPI contribution to 
	the total energy, the standard Hamiltonian for a coupled-EP system can be 
	computed for all semiconductors and insulators.}	

	We report the results of the computations performed for the primitive and 
	supercells of the carbon polytypes, C-dia and C-hex (Lonsdaleite). All 
	computations were performed using the ABINIT 
	software package \cite{Gonze2009, Gonze2016, Abinit}, initally with 
	v.8.8.10 and later (upon referee's suggestion) with the recent 
	version, v.10.2.3. Importantly, the TDepES module \cite{Abinit} can be used 
	with primitive and supercells, as discussed later. The ONCV 
	\cite{ONCVHamann,ONCVPseudodojo} pseudopotential with PBE \cite{PerdewPBE} 
	exchange-correlation functional was used. See Supp. Info. II, III for 
	computational details. We have benchmarked our results for MgO 
	with those of Nery et al. \cite{Nery2018} (Supp. Info. IV, V). 

	Our studies on SCs were motivated by the large energy difference between 
	the PCs of C-dia-2p and C-hex-4p. From these studies, the main theoretical 
	problem has been identified, which is also is seen in the 
	zero-point 	renormalization (ZPR) of the valence band maxima 
	(VBM). Hence, we discuss  the latter first and the main results 
	(EPI contributions to the total energies) alter.

	\begin{table}[h]
		\caption{\label{table1}%
			The ZPR shift ($\Delta \epsilon_{\textrm{ep}}$) at 
			VBM($\Gamma$) for carbon polytypes for primitive and 
			supercells (from (v8.8.10)). The DW ($\Delta 
			\epsilon_{\textrm{ep}}^{\textrm{DW}}$), 
			FM ($\Delta \epsilon_{\textrm{ep}}^{\textrm{FM}}$) and its 
			components, $\Delta 	
			\epsilon_{\textrm{occ}}^{\textrm{FM}}$ and $\Delta 
			\epsilon_{\textrm{unocc}}^{\textrm{FM}}$, are reported. For the 12 
			at/u.c. limited results are reported due to computational 
			constraints. All energies are in meV.} 	
	\begin{tabular}{p{2.3 cm} c c c c c}
		\hline\hline
			Structure & $\Delta \epsilon_{\textrm{ep}}$ & 
			$\Delta 
			\epsilon_{\textrm{ep}}^{\textrm{DW}}$ &$\Delta 
			\epsilon_{\textrm{ep}}^{\textrm{FM}}$ &$\Delta 
			\epsilon_{\textrm{occ}}^{\textrm{FM}}$&$\Delta 
			\epsilon_{\textrm{unocc}}^{\textrm{FM}}$\\
			\hline
			\textbf{C-Diamond}\\
			\hline
			C-dia-2p (prim)\newline (2 at/u.c.)      
			& 142.1 & 1736.8 & -1594.8 & 373.5 & -1968.3 \\ 
			\hline
			C-dia-6 \newline (6 at/u.c.)  
			& 142.1 & 1735.5 & -1593.4 & -69 & -1524.4 \\ 
			\hline
			C-dia-8 \newline (8 at/u.c.)
			& 142.1 &1738.3&-1596.2&-87.6 &-1508.6\\ 
			\hline
			C-dia-12 \newline (12 at/u.c.) 
			& --- &---&---&-130.1 &---\\ 
			\hline\hline
			\textbf{C-Hexagonal}\\
			\hline
			C-hex-4p (prim) \newline (4 at/u.c.)      
			& 117.2 & 1812.8 & -1695.5 & 130.4 & -1825.9 \\ 
			\hline
			C-hex-8 \newline (8 at/u.c.)  
			& 117.5 & 1827.8 & -1710.2 & 83.2 & -1793.4 \\ 
			\hline
			C-hex-12\newline (12 at/u.c.)
			& --- &--- &---& 82.1 &---\\ 
			\hline\hline
		\end{tabular}
	\end{table}

	Table \ref{table1} shows that the ZPR at VBM and its components, DW and FM 
	terms, are identical for the C-dia-2p (PC), C-dia-6 (SC) and C-dia-8 (SC). 
	However, the partial FM terms, $\Delta 
	\epsilon_{\textrm{occ}}^{\textrm{FM}}$ and $\Delta 
	\epsilon_{\textrm{unocc}}^{\textrm{FM}}$, present in the Allen's 
	equation, differ drastically between the PC and SCs. 
	
	For C-hex-4p (PC) and C-hex-8 (SC), the ZPR at VBM are identical and the DW 
	and FM terms differ marginally. However, both $\Delta 
	\epsilon_{\textrm{occ}}^{\textrm{FM}}$ and  $\Delta 
	\epsilon_{\textrm{unocc}}^{\textrm{FM}}$ vary significantly, 
	consistent with the results for C-dia unit cells. 

	In Table \ref{table1}, the FM term, $\Delta 
	\epsilon_{\textrm{ep}}^{\textrm{FM}}$, is summed over infinite number of 
	states ($N_{\textrm{tot}}$) and is independent of unit cell size. However, 
	the partial-FM term, $\Delta \epsilon_{\textrm{occ}}^{\textrm{FM}}$, is 
	summed over valence states ($N_{\textrm{val}}$), and is 
	unit-cell-size dependent. Because this term appears only in the EPI 
	contribution to the whole crystal, it implies that only the EPI 
	contribution to the total energy (and not to the individual eigenstates) 
	will be unit-cell-size dependent.

	We next discuss the EPI contribution to the total energy, and its 
	components, for the primitive and supercells of C-dia and C-hex reported in 
	Table \ref{table2}. These results are similar to those expected from the 
	above discussion. (In ABINIT v10.2.3, the computational cost increases 
	significantly and hence, the emphasis was mostly on computing the $\Delta 
	E_{\textrm{occ}}^{\textrm{FM,tot}}$ term, which causes theoretical 
	problems.)

		\begin{table}[h]
	\caption{\label{table2}%
		The EPI contribution to the total energy ($\Delta  
		E_{\textrm{ep}}^{\textrm{Tot}}$) and its components, EPI contribution 
		to 	the band-structure energy ($\Delta 		
		E^{\textrm{bs}}_{\textrm{ep}}$) and the total FM-occupied contribution 
		($\Delta E_{\textrm{tot}}^{\textrm{FM,occ}}$) for the primitive and 
		supercells of C-dia and C-hex. The computations were 
		made (with tag ieig2rf = 3 for the $M = N_{\textrm{val}}$ runs) using 
		two different 	versions of ABINIT, v8.8.10 and v10.2.3 All 
		energies are in meV/atom.}		
		\begin{tabular}{p{1.9 cm} p{1.9 cm} p{1.9 cm} p{1.9 cm}}
		\hline\hline
		Structure & $\Delta E^{\textrm{bs}}_{\textrm{ep}}$  & $\Delta 
		E_{\textrm{tot}}^{\textrm{FM,occ}}$&$\Delta  
		E_{\textrm{ep}}^{\textrm{Tot}}$\\
		\hline
		\multicolumn{4}{c}{ \textit{ABINIT$-$v8.8.10}}\\ \hline
		\textbf{C-Dia}   \\ \hline
		C-dia-2p      
		& -46.4 & 138.7 & -185.1  \\ \hline
		C-dia-6  
		& -63.1 & -740.3 & 677.2  \\ \hline
		C-dia-8
		& -45.04 & -882.8 &  837.8   \\ \hline
		C-dia-12  
		& --- &-1158.6 & --- \\ \hline \hline
		\textbf{C-Hex}   \\ \hline
		C-hex-4p      
		& -2.75 & -466.68 & 463.93  \\ \hline
		C-hex-8  
		& -23.34 & -883.1 & 859.8 \\ \hline
		C-hex-12
		& ---&-1117.5 &---\\ \hline  \hline 
		
		\multicolumn{4}{c}{ \textit{ABINIT$-$v10.2.3}}\\ \hline
		\textbf{C-Dia}  \\ \hline
		C-dia-2p      
		& -47.14 & -195.3 & 148.2  \\ \hline
		C-dia-4  
		& -31.38 & -523.9 & 492.5  \\ \hline
		C-dia-8
		& --- &-888.2 & --- \\ \hline
		C-dia-12  
		& --- & -1155.6 & --- \\ \hline \hline
		\textbf{C-Hex}   \\ \hline
		C-hex-4p      
		& -14.5 & -516.1 & 501.6  \\ \hline
		C-hex-8  
		& --- & -909.2 & --- \\ \hline
		C-hex-12
		& ---& -1147.5 &---\\ \hline \hline
	\end{tabular}
\end{table}

	In Table \ref{table2}, the EPI contribution to the band-structure energy, 
	$\Delta E^{\textrm{ep}}_{\textrm{bs}}$, is similar, in both versions, for 
	the PCs and SCs for each polytype. In principle, $\Delta 
	E^{\textrm{ep}}_{\textrm{bs}}$ should be identical for PCs and SCs. 
	One possible reason for the slight variation is that the total number of 
	bands, $N_{\textrm{tot}}$, was optimized for only the PCs. For the SCs, for 
	computational reasons, we used $N_{\textrm{tot}}$ to be the number of 
	valence bands plus conduction bands with energies 10-12 eV above the 
	conduction band minima (CBM) \cite{Nery2018}.  

	We next discuss the partial-FM term for \textit{each polytype}. Table 
	\ref{table2} shows that$\Delta E_{\textrm{tot}}^{\textrm{FM,occ}}$
	varies with the unit cell size and has not reached convergence. In 
	particular, $\Delta E_{\textrm{tot}}^{\textrm{FM,occ}}$ varies drastically 
	by $\sim$ 1 eV/atom between C-dia-2p (PC) and C-dia-12 SC. Similar 
	results are obtained for the C-hex PC and SCs, where $\Delta 
	E_{\textrm{tot}}^{\textrm{FM,occ}}$ differs by $\sim$ 0.6 eV/atom meV/atom  
	between C-hex-4p (PC) and C-hex-12 SC. Thus, the unit cell dependence of  
 	$\Delta \epsilon_{\textrm{occ}}^{\textrm{FM}}$ (Table \ref{table1}) and $ 
 	\Delta E_{\textrm{tot}}^{\textrm{FM,occ}}$ is likely to be a universal 
 	feature in all crystalline materials. 	

 	We now compare the partial-FM term, $\Delta 
 	E_{\textrm{tot}}^{\textrm{FM,occ}}$, for the 
 	PCs of C-dia and C-hex. For C-dia-2p and C-hex-4p, the $ 
 	\Delta E_{\textrm{tot}}^{\textrm{FM,occ}}$ term  differs 
 	by O($10^2$) meV/atom. The C-dia and C-hex structures 
 	are analogous to the zincblende and wurtzite structures respectively and 
 	vary only in the orientation of the $sp^3$-bonded tetrahedra 
 	\cite{Zunger1992zinc, Bechstedt2002properties, Bechstedt1994electronic}. 
 	Thus, a large energy difference of O($10^2$) meV/atom is doubtful. This 
 	motivated the EPI studies for various SCs. 

	Table \ref{table2} shows that  partial-FM term, $ 
 	\Delta E_{\textrm{tot}}^{\textrm{FM,occ}}$, differs marginally between the 
 	C-dia and C-hex supercells, when the number of atoms and 
 	hence, valence bands are identical. This is a much more reasonable and 
 	acceptable energy difference given their structural similarities.

	The partial-FM term, $\Delta E_{\textrm{tot}}^{\textrm{FM,occ}}$, in Table 
	\ref{table2} was computed in the $N_{\textrm{val}}$ runs, with the variable 
	ieig2rf = 3, (DFPT method). Similar trends are observed with the 
	variable ieig2rf = 2, (Allen-Cardona method). In particular, $ 
 	\Delta E_{\textrm{tot}}^{\textrm{FM,occ}}$ differs by O($10^2$) meV/atom 
 	between C-dia-2p and C-hex-4p. (Supp. Info. VI). Thus, the energy 
 	differences between the PCs,  C-dia-2p and C-hex-4p, are unacceptable in 
 	both sum-over-states methods.

	Therefore, we report the EPI contributions for the  C-dia-4 (SC) and 
	C-hex-4p (PC) structures, the smallest unit cells with identical number of 
	atoms and valence bands. For the DFT and ZPVE contributions, we report the 
	per-atom values for the C-dia-2p and C-hex-4p PCs, because they are 
	unit-cell size independent.

\begin{table}[h]
	\caption{\label{table3}%
		The various contributions to the standard Hamiltonian of a coupled-EP 
		system, Eq. \ref{Std-Hamiltonian-EPI}, at 0 K, for C-dia and C-hex 
		polytypes. The DFT and ZPVE values are for the PCs while the EPI 
		values are  for C-dia-4 and C-hex-4p unit cells from Table 
		\ref{table2}. All energies are in eV/atom.}		
		\begin{tabular}{p{1.9 cm} p{1.9 cm} p{1.9 cm} p{1.9 cm}}
		\hline\hline
		Structure & $E_{DFT}$  & $E_{ZPVE}$ &$\Delta  
		E_{\textrm{ep}}^{\textrm{Tot}}$ \\
		\hline
		\textbf{C-Dia}\\ \hline
		C-dia-4
		& -163.780& 0.179 & 0.493\\ \hline
		\textbf{C-Hex}\\ \hline
		C-hex-4p  
		& -163.756 & 0.178 & 0.502 \\ \hline
		\textbf{Relative Stability}
		& -0.024 & 0.001 & -0.009 \\ 
		\hline\hline
		\end{tabular}
\end{table}

	\textit{In Table \ref{table3} we report, for the first time, the 
	computation	of the standard Hamiltonian of a coupled-EP system, at 0 K.} 
	The DFT relative stability (24 meV/atom) of C-dia is similar to reported 
	values \cite{Needs2015low}. The ZPVE values are consistent with literature 
	values \cite{Scheffler2018performance, Grochala2014diamond}. \textit{The 
	EPI contribution to the total energy is reported here for the first time.}
 	(The EPI contribution to the phonon energy, not included in 
 	Eq. \ref{Std-Hamiltonian-EPI}, is already included explicitly 
 	in DFPT and implicitly in FDM methods \cite{Allen2020, Giustino2017, 
 	Varma2022}).  Table \ref{table3} shows that the final stability of C-dia-8 
 	over C-hex-8, is 32 meV/atom. It also shows that the EPI term contributes 
 	more than ZPVE to energy differences and must be included in all \textit{ab 
 	initio} total energy studies. 

	As discussed elsewhere \cite{Varma2022}, to include the EPI contributions 
	for large band-gap semiconductors and insulators at finite temperatures, it 
	is only necessary to add the $\Delta E_{\textrm{EP}} (V,0)$ term (Eq. 
	\ref{EPI-zero-K-Allen}) to the the Quasi-Harmonic Approximation (QHA) 
	\cite{Nath2016, Togo2015, Allen2020} expression.  \textit{Including EPI 
	contributions, $\Delta E_{\textrm{EP}} (V,0)$, gives a more accurate free 
	energy at finite temperatures than the QHA.}

	Due to this serious implication of our results, viz. unit cell size 
	dependent total energy due to the partial-FM term, we reiterate the 
	robustness of the entire 	procedure. Firstly, the partial-FM term is 
	derived by Fr{\"o}hlich \cite{Frohlich1950theory}, Fan \cite{Fan1951} and 
	Allen \cite{Allen2022erratum} in independent studies separated by seven 
	decades, which confirm the validity of Eq. \ref{EPI-zero-K-Allen}. 

	Secondly, it is clear from Eq.1 to Eq.5 of Ponc{\'e} \textit{et al.} 
	\cite{Ponce2015},  that the EPI contribution, $\Delta 
	\epsilon_{n\textbf{k}}(0)$, is computed for a supercell with $N_{BvK}$ 
	primitive cells. The TDepES module \cite{Abinit} implements Eq.15 to Eq.17 
	of Ponc{\'e} \textit{et al.} \cite{Ponce2015} where i) the phonon 
	contributions are summed over 3N phonon branches; N being the number of 
	atoms and ii) the derivative with respect to atomic positions and the e-p 
	matrix elements are for supercells with $N_{BvK}$ primitive cells. Clearly, 
	the TDepES module is valid for any unit cell with N atoms. This is an 
	explicit approach in contrast to the use of scaling factors for supercells 
	\cite{Giustino2017}.

	The ZPR values for the PC and SCs are identical and size-consistent as seen 
    from the first three columns of Table \ref{table1}. It is clear that the 
    TDepES module \cite{Abinit} does not contain any deficiencies. 
    Thus, the size dependence of the partial-FM term can only be due to 
    summation over limited number of states ($N_{\textrm{val}}$) instead of 
    infinite number of states ($N_{\textrm{tot}}$), because all other 
    parameters are identical. 

	It is clear that there are no discernible weaknesses in either Eq. 
	\ref{EPI-zero-K-Allen} or in our computations. 

	From a theoretical perspective, in second-order perturbation theory, the 
	difference in the EPI (or FM) contributions to the individual eigenstates 
	and to the whole crystal is clear \cite{Fan1951, Allen2022erratum, 
	Pujari2024}. For a single eigenstate, the summation is over infinite 
	number of states and the EPI (FM) contribution is consistent with 
	lattice periodicity. Whereas, for the whole crystal, an additional 
	partial-FM-occ term, $\Delta\epsilon_{n\textbf{k}}^{\textrm{FM,occ}}$, is 
	present, where the summation is over limited (occupied) states due to the 
	anti-symmetrical nature of the crystal wavefunction, or the Pauli principle 
	\cite{Fan1951, Pujari2024}. This results in unit-cell-size dependent 
	$\Delta E_{\textrm{tot}}^{\textrm{FM,occ}}$ contribution to the per-atom 
	total energy. Therefore, for the whole crystal, the per-atom total energy 
	is unit cell size dependent, even though the crystal is periodic. This is 
	contrary to the understanding \cite{Ashcroft1976solid, 
	Martin2020electronic} of Quantum Mechanics applied to periodic solids and 
	is highly unsatisfactory. 

	Importantly, in the discussion of the partial-FM term (Eq.4) in his 
	original (1951) paper,  Fan \cite{Fan1951} suggested that second-order 
	perturbation theory applied to the whole crystal is of questionable 
	validity, without specifying any reason. Our results support Fan's 
	suggestion for the reason that the partial-FM term, $\Delta 
	E_{\textrm{tot}}^{\textrm{FM,occ}}$, makes the per-atom total energy  
	unit-cell-size dependent. Further, as an alternative, Fan \cite{Fan1951}  
	proposed an approximation, viz. the EPI(FM)	contribution to the 
	band-structure energy is the EPI contribution to the total energy. This 
	implies that in  Eq. \ref{EPI-zero-K-Allen-rewritten},  only the first 
	term, $\Delta E^{\textrm{ep}}_{\textrm{bs}}$, is considered and the 
	second term, $\Delta E_{\textrm{tot}}^{\textrm{FM,occ}}$, is ignored. 
	The neglect of the partial-FM term makes this approximation consistent 
	with lattice periodicity (Tables \ref{table1}, \ref{table2}). However, this 
	approximation, $\Delta E^{\textrm{ep}}_{\textrm{bs}}$ as the 
	sole EPI contribution to the total energy, (independently proposed in our 
	earlier work \cite{Varma2022}) is theoretically invalid as Allen 	
	\cite{Allen2022EPIBS} has shown that it can be derived from the free-energy 
	expression for 	non-interacting quasi-particles, whereas EPI implies 
	interacting quasi-particles (Supp Info VII). Thus, there is currently no 
	theoretically  satisfactory expression for the EPI contribution to the 
	total energy, the $V_{ep}$ term  in the Standard Hamiltonian, Eq. 
	\ref{Std-Hamiltonian-EPI}.
	
	Thus, Fan's suggestion and our results lead to a new fundamental problem in 
	condensed matter physics, viz. second-order perturbation theory is likely 
	invalid for whole crystals. It is essential to resolve this problem, 
	especially, because it causes the Standard Hamiltonian, Eq. 
	\ref{Std-Hamiltonian-EPI}, with its importance for EPI and total energy 
	studies, to be on questionable theoretical foundations.

	In conclusion, for the first time, we have computed the standard 
	Hamiltonian of a coupled EP system, by computing the EPI contribution to 
	the total energy.  This gives the most accurate \textit{ab initio} total 
	energy till date. However, our results show that the  per-atom total energy 
	depends on unit cell size, even though the lattice is periodic. This 
	is due to the partial-FM term that arises from the anti-symmetric nature 
	of 	the crystal wavefunction or the Pauli principle. Due to this, only the 
	energy differences between polymorphs with identical number of atoms per 
	unit cell are meaningful. Thus, our results raise serious theoretical 
	questions.  Our results support Fan's suggestion that second-order 
	perturbation theory is of questionable validity for whole crystals, for the 
	reason that the partial-FM term makes the per-atom total energy unit cell 
	size dependent. Fan's suggestion and our results lead to a new fundamental 
	problem in condensed matter physics, viz. second-order perturbation theory 
	is likely invalid for whole crystals, an entire class. It is essential to 
	resolve this problem as it causes the Standard Hamiltonian, the starting 
	point of EPI studies and also the most accurate \textit{ab initio} total 
	energy, to be on questionable theoretical foundations. 
\\

\begin{acknowledgments}
We thank anonymous referees for their comments. We thank Prof. P. B. Allen for 
sharing a corrected version of Eq. \ref{EPI-zero-K-Allen}. We thank Prof. 
Sumiran Pujari for helpful discussions. We thank the ``Spacetime'' HPC facility 
of IIT Bombay for computational support.
\end{acknowledgments}

	\bibliography{references.bib}

\end{document}